\documentstyle[12pt]{article}
\def\lo{\langle 0 |}
\def\ro{ | 0 \rangle }

\def\gmmu{\gamma _{\mu}}

\def\atop{ \frac{ \alpha_{s}}{8 \pi} G_{\mu \nu}
 \tilde{G}_{\mu \nu} }

\def\gmf{\gamma _{5}}

\def\la{\langle }
\def\ra{ \rangle }

\def\er{ | \eta' \rangle}

\newcommand{\beq}{\begin{equation}}
\newcommand{\eeq}{\end{equation}}
\newcommand{\bea}{\begin{eqnarray}}
\newcommand{\eea}{\end{eqnarray}}

\begin{document}
                                        \begin{titlepage}
\begin{flushright}
hep-ph/9707286
\end{flushright}
\vskip1.8cm
\begin{center}
{\LARGE
 On Topological Susceptibility, Vacuum Energy    \\
\vskip1.0cm
 and Theta Dependence in Gluodynamics 
}         
\vskip1.5cm
 {\Large Igor~Halperin} 
and 
{\Large Ariel~Zhitnitsky}
\vskip0.5cm
        Physics and Astronomy Department \\
        University of British Columbia \\
        6224 Agricultural Road, Vancouver, BC V6T 1Z1, Canada \\ 
        {\small e-mail: 
higor@physics.ubc.ca \\
arz@physics.ubc.ca }\\
\vskip1.5cm
{\Large Abstract:\\}
\end{center}
\parbox[t]{\textwidth}{ 
We suggest that the topological 
susceptibility in gluodynamics can be found in terms 
of the gluon condensate using renormalizability
and heavy fermion representation of the anomaly.
Analogous relations can be also obtained for
other zero momentum correlation functions involving the 
topological density operator. 
Using these relations, we find the $ \theta $ dependence  
of the condensates $ \la GG \ra $, $ \la G \tilde{G} \ra $ and of the
partition function for small $ \theta $ and an arbitrary number of colors.
 }

\vspace{1.0cm}

                                                \end{titlepage}

\section{Introduction}

The importance of exact results in QCD or pure 
Yang-Mills (YM) theory is hard to 
overestimate. There is as yet no thorough understanding of their 
nonperturbative aspects which are of crucial importance in
most instances where strong interactions are involved. In these 
circumstances, 
any exact model-independent statements (theorems) about 
nonperturbative properties of QCD or YM theory become especially
valuable. It turns out that some exact results can be obtained 
even in the  
absence of detailed knowledge of the confinement and chiral 
symmetry breaking mechanisms. These theorems will survive 
any further development of the theory. On the other hand, 
they serve as a test case for any nonperturbative
models and confinement scenarios, which should respect 
them.

Inspired by a recent impressive breakthrough \cite{SUSY}
in supersymmetric (SUSY) theories (see e.g. \cite{Shifman} 
for a review), one may wish to address the 
issue of available exact results for ordinary, non-supersymmetric
QCD or YM theory which is not considered as a softly broken
SUSY model, but is rather taken on its own.
Only a few of them are known so far.
One well known example is provided by the so-called 
t'Hooft consistency condition \cite{t'Hooft} which ensures the
anomaly matching in terms of composite and building block 
particles. Another series of theorems is based on strict 
inequalities
which yield a number of qualitative, but rigorous 
statements for QCD \cite{ineq}.   
Two more classes of exact results are given by anomalous 
Ward identities \cite{Crewther} and low energy theorems based 
on a combination of the Ward identities technique with 
renormalizability arguments \cite{NSVZ}. The power of the latter 
method is that it allows for exact evaluation of some 
nonperturbative
observables in QCD, see e.g. \cite{Shif} for a 
review. Among the variety of results, obtained within this 
approach, exact calculations of zero
momentum correlation functions of scalar gluon currents  
(to be discussed below) are of special interest for our purposes.
 
In this paper we address zero momentum
correlation   
functions of the topological density operator
$ \alpha_s /(8 \pi) G \tilde{G} $ with itself 
and the scalar gluon current $ \alpha_{s}/ (8 \pi) G G  
 $,
in pure YM theory (gluodynamics)\footnote{
We would like to recall that in QCD all zero momentum 
correlation functions of $ G \tilde{G} $ are suppressed 
in the chiral limit by the light quark masses.}.
We will make an attempt to find these correlation functions
in terms of the gluon condensates  
 $ \la GG \ra $, $ \la G \tilde{G} \ra $
without resorting to any particular approximation,
but rather appealing to the general properties of the theory
such as renormalizability and dimensional transmutation.
In particular, we will argue that the topological susceptibility 
in gluodynamics can be found in terms of the gluon condensate
within a particular regularization scheme.
Although our arguments will be somewhat heuristic, there is a   
hope that our final Eq.(\ref{16}) is correct. This hope is supported
by the fact that the same relation can be also obtained along
with a very different line of reasoning \cite{HZ1} which implies 
the same regularization scheme.

Our interest in the correlation functions 
of the above type is mostly related to the 
problem of the $ \theta $ dependence \cite{Jac},
which has been long discussed in the literature practically 
since the discovery of instantons \cite{Pol}. 
We note that little is known about the $ \theta$ dependence 
in YM theory.  Prior to a recent 
renewed interest in these issues in supersymmetric theories 
\cite{SV2,SUSY,Shifman}, essentially all what was known about the 
$ \theta $ dependence in YM theory was the fact that 
the physics should depend on $ \theta $ through the combination
$ \theta /N_c $. This conclusion follows from the analysis
of the famous Witten-Veneziano
mass relation for the $ \eta' $ \cite{Wit} obtained
within the large $ N_c $ 
approach. After the works of Witten and Veneziano, the 
discussion of the $ \theta $ dependence in the literature has 
mostly switched to field theoretical
models other than gluodynamics. One of the purposes of this letter 
is    
to find the $ \theta $ dependence of lowest dimensional
condensates and the partition function in YM theory for 
arbitrary number of colors $ N_c $ and any 
value of $ \theta  \ll \pi $.

There exist many different reasons 
to study the correlation function of the above type
and the $ \theta $ dependence of the lowest dimensional 
condensates.
Here we 
would like to mention a few of them. 
First, such relations are needed for the construction of a 
low energy effective Lagrangian \cite{HZ} which can be used to 
study the vacuum structure of the theory. 
Such an effective Lagrangian allows one to find the $ \theta $ 
dependence
for all (and not only for small) values 
of $ \theta $. Secondly, the latter results for the $ \theta $ 
dependence in YM theory and QCD can be used for the 
axion physics \cite{axion}. Finally, these results 
may be of interest for the lattice studies \cite{Lat} and
nonperturbative 
models of the YM vacuum including, in particular, 
instanton liquid models \cite{Shur}. 

The starting point of our method is the well known relation
\beq
\label{1}
\frac{\alpha_s}{8 \pi} G_{\mu \nu}^a \tilde{G}_{\mu \nu}^a 
= - \lim_{ M \rightarrow \infty}  \la M \bar{ \Psi} i \gmf \Psi 
\ra_{A} \; , 
\eeq
 where $ \Psi $ stands for a fermion field of mass $ M $. The 
meaning of this formula is that the topological density 
operator is {\it equivalent} to the fermion bilinear (\ref{1}) 
in the external gluon field in the 
infinite mass limit\footnote{Note that a different relation 
of the topological density with {\it light} flavour fermion 
bilinears has been used starting from \cite{Bal} in the 
study of the 
$ \theta$-dependence in QCD by current algebra methods.}.   
This
observation determines our whole approach to the 
problem. For our purposes, we consider gluodynamics as the
low energy limit of YM theory coupled to a heavy fermion 
field\footnote{In what 
follows, we use the  
terms gluodynamics and YM theory interchangeably.}.
The fermion mass is assumed to be much larger than the 
inverse confinement radius, which ensures that this heavy fermion
does not affect infrared properties of the theory. 
As its mass can be made arbitrarily large, all calculations are under 
theoretical control, and
no problem can arise with this definition. 
On the other hand, the inclusion of a heavy 
fermion requires a regularization on a much higher ultraviolet
scale. For the latter, we will assume the Pauli-Villars 
regulators. We will therefore work in the double limit 
$ \Lambda_{YM} \ll M \ll M_R $, where $ M $ is the mass of the 
auxiliary fermion, $ M_R $ is the Pauli-Villars fermion mass, and
$ \Lambda_{YM} $ stands for the dimensional 
transmutation parameter in gluodynamics.
We will argue that a combination of the Ward identities technique 
with the fermion representation of the anomaly, 
Eq.(\ref{1}), and its analog for the scalar case
together with renormalizability
arguments allow one to find 
zero-momentum correlation
functions of the topological density in gluodynamics
in terms of the nonperturbative gluon condensates 
$ \la GG \ra $, $ \la G \tilde{G} \ra $, within a particular 
regularization scheme.

Our presentation is organized as follows. In Sect.2 we propose
a method which relates the topological susceptibility in gluodynamics
with the gluon condensate.  
In Sect.3 we combine this result with the
Witten-Veneziano \cite{Wit} approach to the U(1) problem 
at large $ N_c $. The   $ \eta' $ meson mass  
in the limit $ m_q \rightarrow 0 \, , \, 
N_c \rightarrow \infty $
will be expressed in terms of the 
gluon condensate of pure YM theory.
Sect.4 deals with the problem of the 
$ \theta$ dependence in YM theory for small values $\theta \ll \pi $.
Sect.5 contains our conclusions. 


\section{Correlation functions of topological density}
 
In this section we study the zero momentum two-point function
of the topological density (topological susceptibility) in pure
YM theory.
 In what follows we will use the short notation for zero momentum
correlation functions 
\beq
\label{2}
\la A \, , \, B \ra \equiv \lim_{p \rightarrow 0 } 
i \int dx e^{ipx} \lo T \{ A(x) \, B(0) \} \ro 
\eeq
of arbitrary operators $ A(x) $ and $ B(x) $. In this notation, 
we want to evaluate the two-point function :
\beq
\label{3}
P \equiv \la \atop \, , \, \atop \ra \; .
\eeq
As is well known, this correlation function vanishes to 
all order of perturbation theory since the topological density 
operator is a total derivative. Therefore, it seems that explicitly
nonperturbative methods (for example, the instanton approximation) are 
needed to evaluate the two-point
function (\ref{3}).  What will be argued below is that 
this correlation function can be analysed using  
general arguments
appealing to renormalizability of YM theory, together
with the fermion representation of the anomaly, Eq.(\ref{1}). 

 To proceed with our arguments, we would 
like to remind the 
reader how a similar problem of the 
scalar two-point function 
\beq
\label{4}
S \equiv \la \frac{\alpha_s}{12 \pi} G_{\mu \nu}
G_{\mu \nu} \, , \,  
 \frac{\alpha_s}{12 \pi} G_{\mu \nu} G_{\mu \nu} \ra 
\eeq
was addressed by Novikov et. al. (NSVZ) \cite{NSVZ}. These authors 
have derived a fundamental low energy theorem fixing zero momentum 
correlation functions of scalar gluon currents :
\beq
\label{5}
i \int dx \lo T \{ O (x) \; \frac{\alpha_s}{
12 \pi} G^2 (0) 
\} \ro_{YM} = \frac{2 d }{3 b} \la O \ra
\eeq
Here $ O(x) $ is an arbitrary color singlet local
operator of canonical dimension $ d $  
 and $ b = 11/3 N_c $ stands for the 
first coefficient of the $ \beta$-function
in pure Yang-Mills theory\footnote{
The same formula holds in QCD in the 
chiral limit with the substitution
$ b = 11/3 N_c - 2/3 n_f $ where $ n_f $ is a 
number of light flavors.}. As a derivation of  
NSVZ theorem (\ref{5}) is rather simple, for the 
sake of completeness we would like to recall it here.
One starts with a redefinition of the gluon 
field
\beq
\label{6}
\bar{G}_{\mu \nu} \equiv g_0 G_{\mu \nu} \; , 
\eeq
where $ g_0 $ is the bare coupling constant  
defined at the cut-off scale $ M_R $. Then the path
integral representation immediately yields the relation
\beq
\label{7}
i \int dx \lo T \{ O (x) \; \bar{G}^2 (0) 
\} \ro = - \frac{d}{ d(1/4 g_0^2)} \la O \ra \; .
\eeq
On the other hand, the renormalizability and dimensional
transmutation phenomenon in a massless theory (either QCD with 
massless quarks or gluodynamics) ensure that
\beq
\label{8}
\la O \ra = const \; \left[ M_R \exp \left(- \frac{8 \pi^2}{b
g_0^2} \right) \right]^d  
\eeq
with the choice $ b = 11/3 N_c -2/3 n_f $ or $ b =11/3 N_c $, 
respectively.
Finally, performing the differentiation yields the final result 
(\ref{5}). Using the NSVZ theorem (\ref{5}) 
for the diagonal case $ O = (\alpha_s/12 \pi) G^2 $, 
which is proportional to the conformal anomaly $ 
- b \alpha_s/(8 \pi) G^2 $, we obtain
\beq
\label{9}
S =  \la \frac{\alpha_s}{12 \pi} G^2 \, , \,  
 \frac{\alpha_s}{12 \pi} G^2 \ra  = \frac{2}{9 b} \la \frac{
\alpha_s}{ \pi} G^2 \ra \; .
\eeq
One should note that relation (\ref{7}) implies the Wick type of the 
T-product in the correlation function (\ref{9}). Moreover, Eq.(\ref{8})
means that perturbative contributions are subtracted in both sides 
of Eq.(\ref{9}). In this sense, the correlation function (\ref{9}) 
can be obtained by the differentiation of $ \log (Z/Z_{PT}) $
in respect to $ 1/g_{0}^2 $ 
(here $ Z $ and $ Z_{PT} $ stand for the full 
and perturbatively defined partition functions, respectively,
see Eq.(\ref{75}) below)\footnote{ The same result (\ref{9}) was 
obtained in \cite{NSVZ} using canonical methods with 
Pauli-Villars regularization. To one loop order in regulator
fields, it was found that perturbative contributions add the identity
to both sides of Eq.(\ref{9}).}.
 
We would now like to interpret Eq.(\ref{9}) in a  
different way which makes use of
the view of gluodynamics as a low energy limit of 
YM theory coupled to  a heavy   
fermion $ \Psi $ of mass $ M $ 
\footnote{
One should note that both the $ \beta $-function and 
the dimensional parameter $ \Lambda_{YM} $ become different 
when YM theory is coupled to a heavy fermion.
However, the decoupling ensures that the relevant parameters of 
a corresponding low energy effective theory (i.e. pure gluodynamics)
are its own $ \beta $-function and $ \Lambda_{YM} $. The problem 
of matching theories with different color or/and flavor groups
has been discussed more than once in the literature, mostly
in the content of supersymmetric models \cite{SV2,Seib}
.}. 
 An idea similar to what follows  
was first implemented by K\"{u}hn and Zakharov (KZ) to relate 
the proton matrix element of the topological density
$ \la N | G \tilde{G} | N \ra $ to the known quantity $  
 \la N | G^2 | N \ra $ fixed (in the chiral limit) by the conformal
anomaly. A technical trick, which will be used below, was suggested 
in \cite{3} where, in particular, we reproduced by this 
method the KZ relation for the matrix element
$ \la N | G \tilde{G} | N \ra $. In the present work, a similar 
idea is applied to correlation functions of the topological
density.

Let us introduce chiral
projections   
\beq
\label{10}
\Psi_L = \frac{1}{2} ( 1 - \gmf) \Psi  \; , \; 
\Psi_R = \frac{1}{2} ( 1 + \gmf ) \Psi
\eeq 
of the fermion field. We can now use the well known
expansion of the scalar density
\beq
\label{11}
- M \left( \bar{ \Psi}_{L} \Psi_{R} + 
\bar{ \Psi}_{R} \Psi_{L} \right) \rightarrow  
\frac{ \alpha_s}{12 \pi}
G_{\mu \nu} G_{\mu \nu} \; . 
\eeq
Using (\ref{11}), we put Eq.(\ref{9}) in the form
\beq
\label{12}
\la \frac{ \alpha_s}{12 \pi} G^2 \, , \, 
- M \left( \bar{ \Psi}_{L} \Psi_{R} + 
\bar{ \Psi}_{R} \Psi_{L} \right) \ra = \frac{8}{3 b} \la 
- M \left( \bar{ \Psi}_{L} \Psi_{R} + 
\bar{ \Psi}_{R} \Psi_{L} \right) \ra
\eeq
The following observation is crucial for what follows.
Generally, Eq.(\ref{12}) is just equivalent to (\ref{9}), and 
one might therefore think that we would not obtain anything new 
in the course of this identical re-writing. However, one can
note that, as long as the heavy fermion decouples, a ``rotated"
fermion should also do, because the physical content of the 
effective low energy theory 
can not be changed by a redefinition of a heavy fermion field
(see the end of this section for more detail).
In particular, one can consider    
the chiral rotations (change of variables)
\beq
\label{13}
 \bar{\Psi}_{L} \Psi_{R} \rightarrow \exp( i \alpha) 
 \bar{\Psi}_{L} \Psi_{R}  \; , \;  \bar{\Psi}_{R} \Psi_{L}
\rightarrow \exp(- i \alpha) \bar{ \Psi}_{R} \Psi_{L} \; . 
\eeq
Different transformation properties of the two terms in (\ref{12}) 
under rotations (\ref{13}) imply that Eq.(\ref{12}) is equivalent 
to two relations
\beq
\label{k}
\la \frac{ \alpha_s}{12 \pi} G^2 \, , \, 
 M  \bar{ \Psi}_{L} \Psi_{R}  
 \ra = \frac{8}{3 b} \la 
 M  \bar{ \Psi}_{L} \Psi_{R} \ra \; , 
\eeq
\beq
\label{l}
\la \frac{ \alpha_s}{12 \pi} G^2 \, , \,  
M \bar{ \Psi}_{R} \Psi_{L} \ra = \frac{8}{3 b} \la 
 M  \bar{ \Psi}_{R} \Psi_{L} \ra \; .
\eeq
Now we can take the difference of these equations
to extract the combination of interest. One can see that 
in this case the chiral fermion bilinears will combine into the 
pseudoscalar gluon density $ G \tilde{G} $.  
 The relevant 
 expansion reads (cf. Eq.(\ref{1})) 
\beq
\label{14}
- i M \left( \bar{ \Psi}_{L} \Psi_{R} - 
\bar{ \Psi}_{R} \Psi_{L} \right) \rightarrow  
\frac{ \alpha_s}{8 \pi}
G_{\mu \nu} \tilde{G}_{\mu \nu} \; . 
\eeq
Using (\ref{14}), we transform (\ref{k},\ref{l}) into
\beq
\label{15}
\la  \frac{ \alpha_s}{12 \pi} G^2 \, , \,  
\atop \ra = \frac{8}{3 b} \la 
\atop \ra \; . 
\eeq
Normally, both sides of this equation vanish because of the
wrong CP-parity. However, we may think of a ``regularized" theory
with a $ \theta $-term for $ \theta \ll 1 $, where  
Eq.(\ref{15}) would be  
perfectly sensible. In any case, such a ``regularization"
by the $ \theta$-terms is only needed at the intermediate step
(\ref{15}) of our derivation. The final answer will be valid 
for any value of $ \theta $, including $ \theta = 0 $. 
As a double-check of our procedure, one can note that Eq.(\ref{15})
coincides with the general NSVZ formula (\ref{5}) for the 
particular case $ O = (\alpha_s / 8 \pi) G \tilde{G} $. 

Now we proceed similarly with Eq.(\ref{15}). We substitute the 
$ G^2 $ term in the l.h.s. and $ G \tilde{G} $ term in the r.h.s. of 
(\ref{15}) by the heavy quark expansions (\ref{11}) and (\ref{14}),
respectively. Then (\ref{15}) becomes
\beq
\label{15a}
\la \atop \, , \,  
 M \left( \bar{ \Psi}_{L} \Psi_{R} + 
\bar{ \Psi}_{R} \Psi_{L} \right) \ra = i \frac{8}{3 b} \la 
 M \left( \bar{ \Psi}_{L} \Psi_{R} - 
\bar{ \Psi}_{R} \Psi_{L} \right) \ra
\eeq
Using the identity
\[ 
\bar{\Psi}_L \Psi_R - \bar{ \Psi}_R \Psi_L = 
\bar{\Psi}_L \gmf \Psi_R + \bar{ \Psi}_R \gmf \Psi_L 
\] 
and transformation properties under chiral rotations (\ref{13}), we 
see that (\ref{15a}) is again equivalent to two equations 
\beq
\label{15b}
\la \atop \, , \,  
 M  \bar{ \Psi}_{L} \Psi_{R} 
 \ra = i \frac{8}{3 b} \la 
 M \bar{ \Psi}_{L} \gmf \Psi_{R} \ra \; , 
\eeq
\beq
\label{15c}
\la \atop \, , \,  
 M  
\bar{ \Psi}_{R} \Psi_{L} \ra = i \frac{8}{3 b} \la 
 M  
\bar{ \Psi}_{R} \gmf \Psi_{L} \ra \; .
\eeq
It is obvious that the scalar combination of 
Eqs.(\ref{15b},\ref{15c})
exactly reproduces Eq.(\ref{15}) which, as we have seen, is 
a particular version of the original NSVZ theorem (\ref{5}). New 
information is contained in the pseudoscalar part of 
Eqs.(\ref{15b},\ref{15c}). 
Taking the difference of these equations and multiplying
the whole expression by $ i $,
 we finally arrive at the relation
\beq
\label{16}
i \int dx \lo T \left\{ \frac{\alpha_s}{8 \pi} 
G \tilde{G} \, ,  \, 
\frac{\alpha_s}{8 \pi} G \tilde{G} \right\}  \ro =    
 - \frac{2}{9 b} \, \la \frac{
\alpha_s}{ \pi} G^2 \ra \; .
\eeq
The relation obtained is clearly scheme dependent. In fact, 
we have used above the particular scheme in which all
condensates and correlation functions are defined through the 
path integral.
Thus Eq.(\ref{16}) implies the Wick type of the 
T-product, i.e. the two-point function (\ref{16}) is the second 
derivative of $ \log Z $ in respect to $ \theta $. Therefore,
Eq.(\ref{16}) does not contain ultraviolet divergences which are 
present in the factor $ Z_{PT} $ (see Eq.(\ref{75}) below) and 
drop out after the differentiation in $ \theta $. This ensures
that the nonperturbative gluon condensate in Eqs. (\ref{9}) and 
(\ref{16}) is the same quantity. For more comments on the 
scheme dependence of our results, see the end of Sect.4.
We note that after Eq.(\ref{16}) is established, arbitrary 
n-point correlation functions of operators $ G^2 , \tilde{G} G $
can be found using Eqs. (\ref{9}), (\ref{15}) and (\ref{16}) by 
further differentiation in respect to $ 1/g^2 $ and $ \theta $.


Furthermore, as the whole line of reasoning leading
to Eq.(\ref{16}) may appear rather vague, we would now 
like to make further comments justifying all steps made in the 
derivation. 
First, let us  discuss the use 
of the dimensional transmutation arguments (\ref{8}) in our case
where we introduce an auxiliary heavy fermion field. It might 
seem that
in this situation Eq.(\ref{8}) does not hold, since now we have
the mass parameter $ M $ in the theory. Let us, however, notice that 
for a heavy fermion mass $ M $ any dependence on $ M $ in 
formulas like (\ref{8}), 
besides a perturbative redefinition of $ \Lambda_{YM} $, can only 
enter via corrections in powers of  
$ \Lambda_{YM}/ M $ which die off asymptotically as $
M \rightarrow \infty $ (see the comment 
after Eq.(\ref{9})). This is the standard notion of decoupling of the 
heavy fermion.  
Therefore, we are justified in the use 
of Eq.(\ref{8}) in our problem.
 
Secondly, one could object that the rotation (\ref{13})
is not a symmetry, and thus the subsequent manipulations are 
not justified. 
Indeed, the transformation (\ref{13})
is not a symmetry of the theory, and it does change the partition 
function. However, the non-conservation of the current corresponding
to the rotation (\ref{13}) shows up only at the level of $ O(1/M^2) $,
and thus can be neglected in the limit $ M \rightarrow \infty $. 
The latter statement is simply a rephrasing of the textbook fact 
that a heavy quark of mass $ M  \gg \Lambda_{QCD} $ does not 
contribute to the axial 
anomaly. The key difference 
from the case of the light quarks is the cancellation between
the physical and regulator fields at the leading order in $ 1/M^2 $.


Thirdly, 
one could suspect that the substitution (\ref{14}) is very tricky 
due to  the ``double scaling" limit $ \Lambda_{YM} \ll M 
\ll M_R $ which is implied in such replacement. In particular,
one may worry about a region when two densities in 
Eqs.(\ref{15b},\ref{15c}) are close together and essentially form
a new operator of higher dimension which
has different transformation properties. 
However, this small region $x\sim M^{-1}$ 
is under theoretical control, and the operator product expansion
(OPE) can be used to find that the contribution coming from this 
region
would be of order $  M^{-n}$ and is thus negligible.
   
Finally, we would like to discuss the problem of 
higher loop contributions in Eq.(\ref{16}).
In the course of our transformations we have used the one-loop
integration of both the auxiliary and Pauli-Villars fermions. 
Therefore,
our result (\ref{16}) is valid up to higher loop contributions
in correlation functions (\ref{9}) and (\ref{16}). 
Moreover, in the ``double scaling" limit $ \Lambda_{YM} \ll M 
\ll M_R $ which is implied in our calculations, this 
problem is just equivalent to the question
of higher regulator loop contributions to the conformal and axial
anomalies. 
One expects that the inclusion of higher 
regulator loops results in the substitution of the one-loop
anomaly $ - b \alpha_{s}/ (8 \pi) G^2 $ by the full renormalization 
group invariant expression $ \beta(\alpha_{s}) /(4 \alpha_{s}) G^2 $
in zero momentum correlation functions.

\section{Witten-Veneziano formula and gluon condensate}

As an application of the relation (\ref{16}), we can 
consider the famous 
Witten-Veneziano \cite{Wit}
scheme of resolution of the U(1) problem, which is 
based on a large $N_c $ 
approach. We should note that our result (\ref{16})
is valid for arbitrary $ N_c $. However, it is the large $N_c $ 
picture that leads to connections of physical
characteristics of the $ \eta' $ 
with zero momentum correlations functions of 
pseudoscalar 
gluon currents in gluodynamics. As will be shown in this section, 
the low
energy relation (\ref{16}) yields a new   
mass relation for the $ \eta' $ in terms of 
the gluon condensate in gluodynamics.
  
We recall the Witten-Veneziano mass formula for the 
$ \eta' $ \cite{Wit}
\beq
\label{34}
f_{\eta'}^2 m_{\eta'}^2 = - 12 \, \la  
\frac{\alpha_s}{8 \pi} G \tilde{G}  \, , \, 
\frac{\alpha_s}{8 \pi} G \tilde{G} \ra_{YM}  \; , 
\eeq
where $ f_{\eta'} $ is the $ \eta' $ residue 
\beq
\label{35}
 \lo \sum_{q_i = u,d,s} \bar{q}_{i} \gmmu \gmf 
q_{i} \er  = 
i\sqrt{3} f_{\eta'} q_{\mu}  \; , \; 
 \lo n_f \frac{\alpha_s}{4 \pi} G \tilde{G} \er = 
\sqrt{3}f_{\eta'} m_{\eta'}^2  \; 
\eeq
and the two-point function of the topological 
density in (\ref{34})
refers to the pure YM case. It is  
the latter object that is now 
given by Eq.(\ref{16}). We thus arrive at the 
relation which is valid in the double limit $ m_q 
\rightarrow 0  , N_c \rightarrow \infty $ 
\footnote{
Approximate mass relations for the 
$ \eta' $, similar to (\ref{36}), have been repeatedly 
discussed in the literature starting 
from \cite{NSVZ} and \cite{DE}.} 
\beq
\label{36}  
f_{\eta'}^2 \, m_{\eta'}^2 = \frac{8}{ 3 b } \la \frac{ \alpha_s}{
\pi} G^2 \ra_{YM} \; .
\eeq
Our result (\ref{36}) suggests an explanation why the 
correlation function (\ref{34}) is non-zero despite the fact
that it vanishes
to all orders of perturbation theory. In the Witten-Veneziano 
scheme, this fact was assigned to a ghost pole contribution to
(\ref{34}). Our derivation {\it a posteriori} shows that
a non-vanishing value of correlation function (\ref{34}) is due 
to a subtraction constant (gluon condensate)
in the pseudoscalar channels, which is 
proportional to an analogous subtraction constant for the scalar 
channel, and is related to regulator contributions to the conformal
anomaly. The ghost is simply a way to parametrize them.
We note that on a qualitative level, this interpretation 
of the ghost was first discussed in \cite{DE}. 

%

\section{The $ \theta$ dependence in YM theory for $ \theta \ll 
\pi $}

 In this section we will show that the $ \theta$ dependence
of the vacuum energy, topological
density and partition function in YM theory can be found 
{\it exactly} for any $ N_c $ and small $ \theta \ll \pi $ provided
the relation (\ref{16}) takes place.
 More precisely, we will demonstrate that
the Taylor expansions in the vacuum angle $ \theta $ can be re-summed 
exactly for these objects. Thus, formally,
our results hold for any $ \theta $. However, it is believed 
that
the physics is not analytic for $ \theta \sim \pi $. 
This non-analyticity is related to the existence of distinct disconnected
vacua which cross in energy at $ \theta \sim \pi $. The Taylor expansion
in $ \theta $ refers to a state of lowest energy (at small $ \theta $)
out of this set, and thus can not probe other states. As a result, it
can be trusted only for $ \theta \ll \pi $. As will be discussed below,
the existence of additional vacua is crucial to establish
the correct periodicity in $ \theta $ with period $ 2 \pi $.
A corresponding analysis \cite{HZ} reveals that 
indeed Eqs.(\ref{73}),(\ref{74}) (see below) 
correctly describe the $ \theta $ 
dependence for small $ \theta \ll \pi $.
     
Let us first address the 
$\theta$ dependence of the vacuum energy. The latter is 
defined by the $\theta$-vacuum expectation value of the trace
of the momentum-energy tensor :
\bea
\label{71}
E_{vac} &=& - \frac{b}{32} 
\la \frac{ \alpha_s}{ \pi} G^2 \ra_{\theta}  
= - \frac{b}{32} \left( \lo \frac{ \alpha_s}{ \pi} G^2 \ro 
\right.\nonumber \\
& +&  \left. \frac{\theta^2}{ 2!} i^2 \int dx \, dy \, 
\lo T \left\{ \frac{ \alpha_s}{ \pi} G^2 (0) \,  
\frac{ \alpha_s}{ 8 \pi} G \tilde{G} (x) \,   
\frac{ \alpha_s}{ 8 \pi} G \tilde{G} (y) \right\} \ro + 
\ldots \right) \; .
\eea
The three-point correlation function is now 
fixed by the low energy relations
(\ref{5}) and (\ref{16}). Indeed, differentiating Eq.(\ref{15}) in 
respect to $ \theta $, using Eq.(\ref{16}) and setting $ \theta $ to 
zero in a final expression, we obtain
\beq
\label{72}
 i^2 \int dx \, dy \, 
\lo T \left\{ \frac{ \alpha_s}{ \pi} G^2 (0) \,  
\frac{ \alpha_s}{ 8 \pi} G \tilde{G} (x) \,   
\frac{ \alpha_s}{ 8 \pi} G \tilde{G} (y) \right\} \ro =
- \left( \frac{8 }{ 3 b} \right)^2   
\lo \frac{ \alpha_s}{ \pi} G^2 \ro \; .
\eeq
As before, the Wick T-product is implied in Eq.(\ref{72}). We thus 
see that the three-point function (\ref{72}) is expressed in 
terms of the same nonperturbative gluon condensate which enters
Eq.(\ref{9}) and (\ref{16}).    
Repeating the same procedure to all orders in $\theta$ in expansion
(\ref{71}), we arrive at the following result  
for $ \theta \ll \pi $ :
\beq
\label{73}
E_{vac} = \la \theta | - \frac{b \alpha_s}{32 \pi}  G^2
| \theta \ra =    \lo - \frac{b \alpha_s}{32 \pi}  G^2 \ro
\, \cos \left( \frac{ 8}{3 b} \theta \right) \; .
\eeq
The $ \theta $ dependence of the topological density condensate
can be found analogously :
\beq
\label{74}
\la \theta | \frac{ \alpha_s}{ 8 \pi} G \tilde{G} 
 | \theta \ra =  -  \lo \frac{ \alpha_s}{12 \pi}  G^2 \ro
\, \sin \left( \frac{ 8}{3 b} \theta \right) \; .
\eeq
Similarly, we can calculate the $ \theta $ dependence of the 
partition function $ Z_{\theta} $ of YM theory. Since $ \log Z_{\theta}
$ generates connected n-point functions of the topological
density, resumming the expansion of  $ \log Z_{\theta}
$, we obtain 
\beq
\label{74a}
\log Z_{\theta} = const + i \frac{b}{32} 
V \lo \frac{ \alpha_s}{ \pi}
G^2 \ro \, \cos \left( \frac{ 8}{3 b} \theta \right) \; ,
\eeq
where $ V $ is the four-volume. A constant in Eq.(\ref{74a}) 
can be found
from the relation \cite{Diak}
\beq
\label{74b}
\log Z_{0} = \log Z_{PT} + i \frac{b}{32} V \lo \frac{ \alpha_s}{ \pi}
G^2 \ro \; , 
\eeq
where $ Z_{PT} $ stands for the perturbatively defined partition 
function. We note that a perturbative contribution to the 
gluon condensate 
is absorbed in the definition of $ \log  Z_{PT} $. We therefore 
obtain
\beq
\label{75}
Z_{\theta} = Z_{PT} (g_{0}^2)  \, \exp \left\{ i \frac{b}{32} V \lo 
\frac{ \alpha_s}{ \pi}
G^2 \ro \, \cos \left( \frac{ 8}{3 b} \theta \right) \right\} \; .
\eeq
One can see that $ \theta $ enters relations (\ref{73}), (\ref{74}) 
and (\ref{75}) in the combination $ \theta / N_c $. Thus, 
our results confirm the long standing conjecture, first arrived
at within the large $ N_c $ picture, that a $ \theta $ dependence 
should come in the combination $ \theta / N_c $ \cite{Wit}.
At the same time, these results seem to suggest a ``wrong" periodicity
in $ \theta $. As was mentioned above, this interpretation of 
Eqs. (\ref{73}), (\ref{74}) 
and (\ref{75}) would be wrong, as they are valid only for small
values of $ \theta \ll \pi $. It can be shown \cite{HZ} that
the correct $ 2 \pi $ periodicity in $ \theta $ is recovered when
additional nondegenerate disconnected vacua, which cannot be seen in the 
naive thermodynamic limit $ V \rightarrow \infty $, are taken into 
account. The correct periodicity in $ \theta $ is the property
of this set of vacua as a whole, while some two of them cross
in energy at $ \theta \sim \pi $ \cite{HZ}.
For small $ \theta \ll \pi $ these additional states 
do not contribute the partition function in the limit   
 $ V \rightarrow \infty $, and correspondingly are not probed
by the Ward identities. Their existence is rather
revealed with an effective Lagrangian approach, see \cite{HZ}.
A corresponding analysis shows that Eqs. (\ref{73}), (\ref{74}) 
and (\ref{75}), as standing for
small values of $ \theta \ll \pi $, are 
not in conflict with general principles 
of analyticity and $2\pi$ periodicity in $ \theta $. 
 
Finally, we would like to comment on the scheme dependence 
of our results. As was mentioned earlier,  
there is 
a specific scheme which is implied in all our formulas. Namely,
we use the path integral definition for  all correlation functions
as the derivatives with respect to appropriate sources
(parameters $1/g_{0}^2$ or $\theta$). This definition 
of zero momentum correlation functions implies the Wick type of the 
T-product symbol.
It also  fixes the rule of subtraction of the  perturbative UV divergent 
contributions to the correlation functions, once such a rule 
is imposed for the gluon condensate. 
In terms of Eq.(\ref{75}), this procedure means that the  
NSVZ low-energy theorem (\ref{9}) is obtained  by differentiation
of the logarithm of 
$ Z_{\theta}/ Z_{PT} $ with respect to $1/g_{0}^2$, see Eq.(\ref{75}). 
Our relation (\ref{16}) and its $n$-point generalizations
are obtained by differentiation
of  the same expression $ \log (Z_{\theta} / Z_{PT}) $ (\ref{75}) 
with 
respect to $\theta$.
It is clear that a change of the subtraction prescription 
in this scheme will generally alter the absolute value of the 
condensate,
but it does not change the general structure of these relations, 
see also the appendix in \cite{HZ} for more detail.

Last, but not least, formula (\ref{75}) implies a 
``hidden'' symmetry which may be thought of as the  
origin of the new low-energy relations similar
to (\ref{16}). This hidden symmetry arises due to
the appearance of the new complex parameter 
$\tau= 1/ g_{0}^2 + i \theta /(12 \pi^2 )$ in Eq.(\ref{75}). Indeed,
Eq.(\ref{75}) implies the relation (for small $ \theta $)
\beq
\label{pq}
 \la - \frac{ b \alpha_s}{ 8 \pi} G^2 \ra_{\theta} = 
const \; Re \; M_{R}^4 \exp \left( - \frac{32 \pi^2}{b g_{0}^2 }
- i \frac{8}{3b} \theta \right)  \equiv 
const \; Re \; M_{R}^4 \exp \left( - \frac{32 \pi^2}{b} \tau 
\right) \; .  
\eeq  
Thus, the 
{\it nonperturbative } vacuum energy depends just on this 
single combination of the coupling constant and $ \theta $ angle.
This is exactly the origin of the new relation (\ref{16})
which together with the NSVZ theorem (\ref{9}),(\ref{15}) 
can be rewritten
in a holomorphic way \cite{HZ}. With our definition
of the {\it nonperturbative} correlators as the derivatives of the 
partition function with respect to the sources $\tau , 
\bar{\tau} $, the
{\it nonperturbative}
vacuum energy in Eq.(\ref{75}) is the only relevant term 
to differentiate.
A non-holomorphic dependence on $g_{0}^2 $ of  the perturbative
part $Z_{PT}$ is irrelevant in our definition for calculation of the non-perturbative
correlation functions. At the same time, the $\theta$ dependence 
appears only
in the non-perturbative part of $Z_{\theta}$. Therefore, the 
origin of our new 
low-energy relations can be understood as the interplay between the exact 
 NSVZ theorem (\ref{9}) and assumption of 
separation of the perturbative and non-perturbative contributions  
in Eq.(\ref{75}).
Once this assumption is made, a new complex structure emerges
due to the pure non-perturbative origin of the $\theta$ parameter 
which combines 
with another parameter $1/g^2$ into the unique complex combination $\tau$.

\section{Conclusions}

Low energy theorems provide a bridge between 
low energy and high energy physics, including phenomena
at the boundary of the ultraviolet cut-off. The fact that 
the axial anomaly can be described in terms of regulator 
fermions was emphasized by Gribov \cite{Gribov} 
(see also \cite{Shif}) long ago. In this paper
we suggested using the connection between the axial 
anomaly term and the regulating Pauli-Villars fermion bilinear, 
given by Eq.(\ref{1}), 
to derive new relations for zero momentum 
correlation functions of the topological density.
Classically, 
the topological
density operator is a total derivative, which 
suggests that its effects can only be treated within 
infrared, explicitly nonperturbative methods such as the 
instanton approach. Yet, one can use another, quantum, definition
of this object, which is given by Eq.(\ref{1}). In contrast
to the (quasi-) classical treatment, none of the extremely
complicated problems of the latter (for example, a compactification
of the infrared boundary, summation over different topological 
classes, instanton interactions, etc.) arise in the quantum 
approach. Instead, the problems are shifted to the analysis 
of the high energy behavior which is fixed by renormalizability.
Therefore, 
we believe that a look at the ``dark", ultraviolet side
of the topological density is very instructive as it can test
nonperturbative infrared methods. An interesting question which 
could be asked 
in reference to Eq.(\ref{1}) is whether this quantum definition
of the topological density can be used in lattice studies 
of QCD and gluodynamics.    
 
New low energy relations, taken together with analogous 
formulas for 
the scalar channel \cite{NSVZ}, enable us to 
evaluate some nonperturbative observables 
in gluodynamics, which seem to be out of the 
reach of other methods,
within the regularization scheme based on the path integral.
In particular, in this scheme we have found 
the relation between the topological susceptibility
in YM theory, which is one of the most important 
characteristics of the YM vacuum, and the 
nonperturbative gluon condensate.
Furthermore, within the same scheme 
we calculated the $\theta$ dependence for
arbitrary $ N_c $ and small $ \theta \ll \pi  $ 
for lowest dimensional vacuum condensates and for the 
partition function in gluodynamics, and 
found that the $ \theta $ parameter always comes in the 
combination $ \theta/ N_c $, as was expected for a long time.
Our results can have various applications which have been already 
mentioned in the introduction.

\end{document}